**A Multi-Modal Sensor Fusion Instrument for Measuring Regional Human Mobility: The Distributed Human Data Engine (DHDE)**

**Authors:**
Amil Khanzada[1] (ORCID: 0000-0002-7065-5390), Takuji Takemoto[2] (ORCID: 0000-0002-1150-2714)

**Affiliations:**
[1] Specially Appointed Assistant Professor, Headquarters for Regional Revitalization, University of Fukui, Japan.
[2] Professor, Headquarters for Regional Revitalization, University of Fukui, Japan.

***Corresponding Author:** Amil Khanzada amil.k@u-fukui.ac.jp

## Abstract

Accurately estimating human mobility in peripheral regional economies presents a fundamental measurement challenge: physical ground-truth sensors are sparse, behavioral intent signals are heterogeneous, and environmental friction introduces systematic bias into demand inference. We introduce the Distributed Human Data Engine (DHDE), a multi-modal sensor fusion architecture that addresses this challenge by integrating physical instrumentation (Edge-AI cameras), digital intent signals (route search impression metrics), behavioral records (90,350 spending records, 97,719 standardized survey responses), and meteorological data across four geographically distributed nodes in Fukui, Japan.

The primary measurement-science contribution is the design, deployment, and cross-node validation of the DHDE as a sparse-sensor compensation instrument: a heterogeneous sensor fusion architecture that anchors non-stationary digital intent signals to concurrent physical ground-truth counts, correcting for systematic bias introduced by meteorological planning friction. The instrument is implemented as an ensemble inference pipeline (Random Forest and Ordinary Least Squares with Newey-West robust inference), calibrated across 397 daily observations and validated by chronological holdout replication across four geographically distinct node types.

The primary OLS specification achieved an in-sample explanatory power of $R^2 = 0.810$ and a chronological out-of-sample predictive performance of $R^2 = 0.683$. Results identify an Under-Vibrancy Paradox where macro-regional visitor satisfaction correlates positively with crowd density (Spearman rank correlation $r_s = +0.150$, $p = 0.002$). We estimate an annual proxy gap of 865,917 intent-implied visits, corresponding to JPY 11.96 billion (USD 72.6 million) in foregone revenue. The validated instrument output provides a metrologically grounded evidence base for resource allocation in peripheral regional economies.

## 1. Introduction

Accurately measuring human mobility in peripheral regional economies presents a fundamental instrumentation challenge. The prevailing literature in urban informatics and tourism management disproportionately focuses on the negative externalities of high visitor density, a phenomenon widely characterized as "overtourism" in primary global destinations such as Kyoto and Venice [1,2]. Conversely, the structural stagnation of secondary or peripheral municipalities remains critically underrepresented. These regions frequently suffer from under-vibrancy: a socio-economic condition defined by sub-optimal spatial utilization, diminished atmospheric vitality, and subsequent economic leakage. In such contexts, the primary threat to regional sustainability is not congestion, but severe planning friction that prevents latent digital demand from converting into physical visitation.

The failure to capture this latent demand is rarely due to insufficient cultural or natural resources. Rather, it is driven by environmental stochasticity and static municipal governance frameworks. To investigate this dynamic, this study focuses on Fukui Prefecture, located in central Japan along the Sea of Japan. Despite its adjacency to hyper-dense global hubs like Kyoto and proximity to Osaka, Fukui remains Japan's least-visited prefecture and is characterized by harsh winter micro-climates, including heavy coastal winds and inland snowfall. In such environments, high digital intent, as observed in search engine queries and map routing requests, often dissipates before physical arrival due to weather uncertainty. This cancellation behavior triggers a negative feedback loop: local merchants preemptively reduce operational hours, exacerbating the perceived emptiness of the destination, which in turn degrades the post-visit satisfaction of the remaining visitor cohort. Addressing this structural failure requires advanced estimation techniques to transition from reactive policy formulation to proactive, data-driven governance. Consequently, we apply machine learning models not merely as predictive tools, but as robust measurement instruments to quantify the exact scale of "planning friction" across the region.

To address this measurement gap, this study introduces the Distributed Human Data Engine (DHDE), a multi-modal measurement instrument and intelligent decision support system (IDSS) that fuses heterogeneous sensor streams into a calibrated, validated estimate of regional human mobility, compensating for the sparse-sensor problem characteristic of peripheral economies. Although the DHDE is novel in the context of regional tourism governance, it draws conceptual inspiration from prior public health research in which large-scale crowdsourced behavioral data were aggregated to generate actionable insights for AI-assisted diagnosis [3]. Adopting an Input-Process-Output (IPO) architecture similar to that used in clinical data collection, the DHDE is generalized here to forecast spatial demand and support socio-economic load balancing.

Building on this framework, we implement a comprehensive DHDE ecosystem across four geographically saturated nodes (coastal, urban transit, mountainous, and scenic corridor) within Fukui Prefecture in Japan's Hokuriku region. The system integrates high-granularity, multimodal data streams, including route search impression data from the Code4Fukui open data repository, hourly micro-climate observations from the Japan Meteorological Agency (JMA), physical ground-truth visitor counts from edge-AI cameras (supplemented by survey proxies), and behavioral metrics from a large-scale tourism survey database comprising 97,719 standardized responses. By applying Kansei (affective) information-science methodologies grounded in regionally developed affective analysis frameworks in Fukui [4], we examine how visitor satisfaction relates to spatial density and micro-level carrying capacities, empirically demonstrating the region's atmospheric resilience and its latent capacity to absorb rerouted visitor flows.

The primary contributions of this research are as follows:

1. **Sensor Fusion Instrument Validation:** We validate the DHDE's measurement fidelity against physical ground-truth counts across 397 daily observations at four geographically distinct node types, establishing in-sample explanatory power of ($R^2 = 0.810$). Chronological holdout replication on an independent 80-day period confirms metrological stability under temporal transfer (out-of-sample $R^2 = 0.683$), demonstrating that the fused sensor signal provides reliable substitution for sparse physical ground-truth coverage.
2. **Identification of the Under-Vibrancy Paradox:** We provide empirical evidence that in structurally weak regional economies, visitor satisfaction is positively correlated with crowd density, fundamentally challenging prevailing overtourism-centric models.
3. **Quantification of Economic Leakage:** We define and quantify an annual "Opportunity Gap" of approximately ¥11.96 billion (USD 72.6 million at ¥164.7/$1) in foregone regional revenue across four sites due to weather-degraded intent-to-arrival conversion rates.
4. **Measurement-Grounded Decision Support:** The DHDE's calibrated proxy gap estimate of ¥11.96 billion in foregone regional revenue provides planners with a metrologically grounded evidence base for resource allocation, demonstrating that validated fused-sensor output directly informs economic planning in sparse-sensor peripheral economies.

## 2. Related Work

The theoretical and methodological foundations of this study reside at the intersection of urban informatics, distributed governance, predictive demand analytics, behavioral economics, and Kansei information science. By reviewing the progression of these domains, we delineate the boundaries of existing research and establish the original scientific contributions of this study.

### 2.1 Urban Informatics and the Smart Tourism Destination

The rapid digitalization of global travel has catalyzed the emergence of the Smart Tourism Destination (STD) paradigm. STDs are defined by their ability to integrate Internet of Things (IoT) infrastructure, cloud computing, and end-user mobile connectivity to enhance destination competitiveness and enrich visitor experiences [5,6]. A vast majority of recent STD literature is heavily weighted toward managing "overtourism," which occurs when visitor density surpasses physical or ecological carrying capacities, thereby degrading the quality of life for residents and the quality of the tourist experience [7,8]. Consequently, the spatial optimization literature prioritizes capacity restrictions, congestion pricing, and de-marketing strategies for hyper-dense global hubs such as Kyoto, Barcelona, and Venice [1,2].

However, this focus creates a distinct theoretical gap regarding peripheral or lower-ranked municipalities. For regions experiencing population decline and structural stagnation, the primary socio-economic threat is not congestion. Existing literature often diagnoses spatial emptiness in these areas as a fundamental lack of attractive resources, prescribing traditional destination marketing as the primary solution [9]. To address this oversight, the present study introduces the novel concept of under-vibrancy. We mathematically frame spatial emptiness as an environmentally induced planning-friction problem that actively degrades the visitor experience and is associated with economic leakage, arguing that an optimal level of crowding is actually a prerequisite for satisfaction in structurally under-visited environments.

## 2.2 Distributed Governance and Human-as-a-Sensor Networks

The concept of utilizing distributed human populations as data-gathering nodes has its origins in participatory sensing and crowdsensing models, where mobile devices act as ubiquitous environmental monitors [10,11]. In recent years, this socio-technical architecture has proven highly effective in rapid crisis management. For example, Author et al. (2025) demonstrated that large-scale crowdsourced biological data could be aggregated using behavioral nudges to train diagnostic artificial intelligence models during the COVID-19 pandemic. This established the viability of an Input-Process-Output architecture where human behavior serves as a real-time sensor network, bypassing the latency of traditional bureaucratic reporting.

While the literature widely supports crowdsensing for disaster response and public health [12], its continuous application for algorithmic governance of regional economic health remains underexplored. Economic governance still largely relies on lagging macroeconomic indicators rather than real-time behavioral signals [13]. This study contributes to the literature by generalizing the Distributed Human Data Engine (DHDE) framework. By shifting its application from biological crisis mitigation to economic prosperity optimization, we demonstrate that human-as-a-sensor networks can power evidence-based policy making (EBPM) in regional tourism. This is consistent with recent developments in multimodal data fusion, which emphasize integrating heterogeneous sensors to build intelligent decision-support systems for real-time urban management [14].

## 2.3 Predictive Analytics and Environmental Friction

Predicting human mobility and tourism demand relies heavily on digital intent signals. Choi and Varian [15] pioneered the use of search engine query data to forecast near-term economic indicators, establishing a method that has since become a standard in tourism demand forecasting [16,17]. Unlike Google Trends, which aggregates search query volume as a population-level index, this study employs route search impressions (RSI): a place-specific navigation intent signal capturing users who actively sought a route to each monitored destination. This distinction is methodologically significant, as RSI represents a higher-specificity, place-anchored behavioral intent measure than aggregated query frequency data. Advanced forecasting models have evolved from simple autoregressive integrated moving-average (ARIMA) techniques to complex machine learning algorithms, including Random Forests and Long Short-Term Memory (LSTM) neural networks [18].

Despite these algorithmic advancements, most predictive models treat exogenous environmental variables, such as precipitation and extreme temperatures, as secondary statistical noise rather than primary causal mechanisms. Day et al. [19] noted that weather is a critical behavioral gatekeeper in tourism, yet few forecasting models dynamically integrate high-resolution meteorological data to quantify specific trip cancellation rates at the micro-destination level. This study advances the methodology of predictive analytics by proposing a novel approach that integrates route search impression-based intent signals with Japan Meteorological Agency (JMA) micro-climate data. This allows us to isolate "planning friction" and estimate a proxy for the economic opportunity gap associated with weather-associated intent-arrival mismatch.

## 2.4 Algorithmic Governance and Digital Nudging

Behavioral economics, particularly Nudge theory as formalized by Thaler and Sunstein [20], proposes that choice architecture can predictably alter human behavior without forbidding any options or significantly changing economic incentives. In public policy, frameworks such as EAST (Easy, Attractive, Social, and Timely) have been utilized to encourage tax compliance, organ donation, and public health adherence [21]. The migration of these concepts into digital interfaces has birthed the concept of "digital nudging" [22], which, when paired with artificial intelligence, enables "algorithmic governance" [23].

In tourism research, nudging has primarily been applied to micro-level sustainability behaviors, such as encouraging towel reuse in hotels or reducing food waste, rather than to macro-level spatial load balancing. Similarly, most tourism routing and recommendation algorithms remain optimized for transit efficiency or congestion avoidance, with limited consideration of experiential or atmospheric conditions at the destination [6,24]. This research extends the literature by introducing an original dual-nudge feedback loop for spatial governance: a closed-loop system in which supply-side nudges alert merchants to adjust operating hours based on forecasted demand, while demand-side nudges dynamically reroute visitors toward weather-resilient locations under conditions of elevated environmental planning friction.

### 2.5 Kansei Information Science in Spatial Management

Kansei Engineering evaluates human psychological and emotional responses to external physical stimuli, translating subjective feelings into quantifiable design parameters [25]. Originally developed for industrial product design, recent advancements in regional informatics have expanded Kansei methodologies to evaluate affective responses to municipal environments and regional branding [26]. For example, [4] utilized semantic differential scales and factor analysis to quantify university students' emotional attachment and regional pride regarding Fukui Prefecture. Their work demonstrated how these internal Kansei evaluations directly influence concrete behaviors, such as employment location choices.

The present study builds upon this regionally grounded Kansei approach by applying rule-based affective keyword analysis to large-scale visitor satisfaction surveys. We extend these Kansei principles from static image evaluation to dynamic spatial capacity management. By correlating crowd density with subjective satisfaction scores, we empirically evaluate the atmospheric resilience of the region's diverse attractions. This establishes a framework to quantify latent carrying capacity, ensuring that future spatial load-balancing strategies can safely redistribute visitor flows without assuming premature congestion.

## 3. Methodology

This study operationalizes the Distributed Human Data Engine (DHDE) as a novel multi-modal measurement instrument and forecasting system to examine planning friction and regional under-vibrancy. The methodology is structured into five phases: multi-source data harmonization, predictive feature engineering, statistical modeling and validation, Kansei information extraction, and economic opportunity quantification. The regional tourism system is treated as a continuous, distributed sensor network capturing the progression from digital intent to physical arrival and post-visit affective evaluation (see Figure 1).

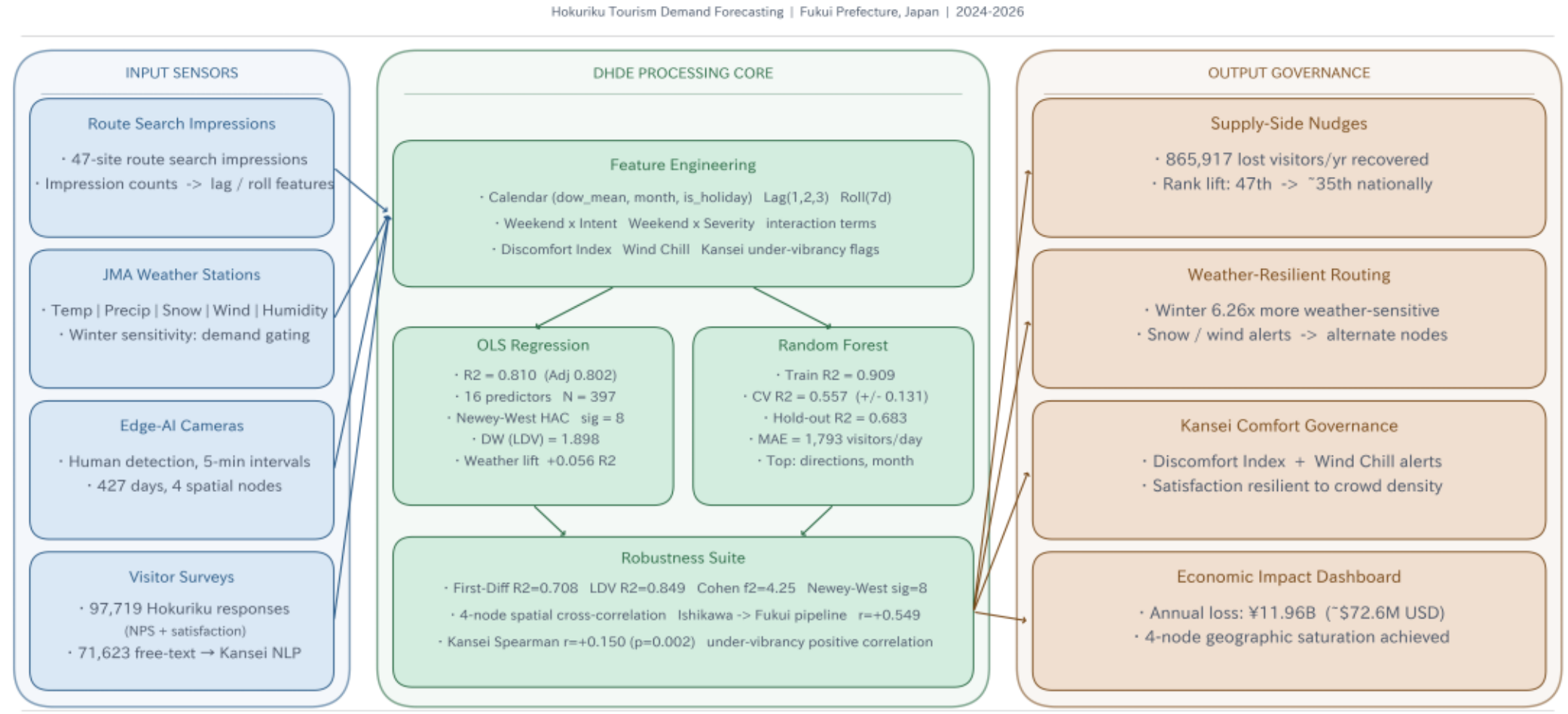

**Figure 1. The Distributed Human Data Engine (DHDE) decision support system architecture.** The framework integrates four heterogeneous sensor modalities into a unified measurement and multi-variable processing core. The architecture utilizes a robust suite of statistical validations to ensure predictive stability as the underlying database scales. The resulting policy intelligence enables targeted planning interventions for regional economic management.

## 3.1 Multi-Source Data Harmonization and Geographic Saturation

The DHDE integrates four heterogeneous data streams synchronized by date and geographic location. The observation period for physical-flow modeling spans 427 usable days over approximately 15 months, from December 20, 2024 to March 10, 2026, after excluding sensor outage intervals. To ensure spatial representativeness across the prefecture's diverse topography, we achieved geographic saturation by selecting four structurally distinct monitoring nodes: Tojinbo as a weather-exposed coastal landmark, Fukui Station as a central urban transit hub, Katsuyama as a mountainous heritage and dinosaur museum node, and the Rainbow Line as a highly seasonal scenic drive corridor. In addition to the four primary governance nodes, the Eiheiji Zen temple was analyzed as a secondary heritage sub-site for survey-based Kansei analysis. This provides a culturally sensitive control case to test the atmospheric resilience of the region's most fragile spiritual landmarks. Conceptually, the system performs knowledge representation by mapping unstructured human affective feedback (Kansei signals) and exogenous environmental constraints into a structured rule base for regional management.

The integrated data streams are defined as follows:

1. **Digital Intent Data:**
   Daily route search impression (RSI) data were obtained from the Fukui Tourism Location Trend Report open data repository (Code for Fukui; see Appendix A.3), which aggregates impression counts from online maps and web search tools for 47 tourism locations across Fukui Prefecture. The daily aggregate of navigation route requests for each tourist destination was isolated as the primary proxy for latent travel demand, representing user commitment beyond passive information search.

2. **Environmental Filter Data:**
High-resolution meteorological observations were obtained from four Japan Meteorological Agency (JMA) stations (Mikuni, Fukui City, Katsuyama, and Mihama). Local observatory data corresponding to each node were used to capture micro-climate variation. Hourly precipitation, temperature, sunshine duration, wind speed, and snow depth were aggregated into daily profiles.

3. **Ground-Truth Physical Data:**
DNN Network Bullet IP Camera U2SM-B units (IP66-rated; on-camera YOLO-based human-shape detection) recorded visitor counts at five-minute intervals for three of the locations. The IP66 ingress protection rating ensures resistance to dust and powerful water jets, providing documented operational robustness against the coastal wind-driven precipitation and heavy snowfall characteristic of Fukui's Sea of Japan climate. Node D (Rainbow Line) utilized a parking gate facial-detection sensor; because this represents a sub-sample of total vehicle occupancy rather than a total site count, it exhibits higher instrumental noise ($R^2 = 0.168$) compared to other nodes. Node C (Katsuyama) relied on a validated survey-response proxy. This proxy was validated by correlating daily survey volumes with ground-truth camera counts at the primary node ($r = 0.564$, $p < 0.001$), confirming that survey response frequency is a reliable indicator of physical density in this regional context. Camera data was aggregated into daily totals, and an automated anomaly detection process removed 17 zero-count days associated with documented sensor outages.

4. **Data Integration and Ethics:**
This study integrates two complementary data streams. Dataset 1 comprises 90,350 raw Fukui-specific records (spanning 576,127 physical CSV lines) utilized for spending analysis. Dataset 2 comprises 97,719 standardized responses from across the Hokuriku region (Fukui, Ishikawa, and Toyama) utilized for Kansei sentiment mining and predictive modeling. All datasets were de-identified and provided as secondary data by the Hokuriku Inbound Tourism DX and Data Consortium in accordance with regional privacy guidelines and institutional ethics standards. This study uses secondary, de-identified administrative data provided by a public-sector consortium; no primary human subjects research was conducted by the authors, and formal IRB review was not required under applicable institutional guidelines.

### 3.2 Predictive Feature Engineering

To model the temporal gap between planning and execution, digital intent variables were lagged from zero to seven days. A seven-day rolling average was also computed to mitigate weekday search volatility.

Environmental variables were treated as exogenous friction factors. To capture non-linear behavioral responses, we constructed a composite Weather Severity Index ranging from 0 to 3. Thresholds were defined using empirically grounded safety and accessibility cutoffs: precipitation exceeding 10 mm/day, wind speeds above 8 m/s, and accumulated snow depth constraining vehicular access, where available. Snow depth data were available only at Node B (Fukui City main observatory); the remaining three nodes rely on precipitation and wind as primary weather friction indicators. Calendar effects were controlled using binary indicators for weekends and Japanese national holidays, alongside interaction terms combining calendar and weather severity.

### 3.3 Statistical Modeling and Robustness Validation

The predictive objective was to estimate the relative contribution of psychological intent and environmental friction to physical arrivals. A multiple Ordinary Least Squares (OLS) regression model served as the primary explanatory framework. Standardized Beta coefficients ($\beta$) were computed to enable direct comparison of effect sizes.

To address time-series dependence, residuals were evaluated using the Durbin-Watson statistic. Identified autocorrelation was corrected using Newey-West heteroskedasticity and autocorrelation consistent estimators. Additional robustness checks included a First-Difference specification and a Lagged Dependent Variable (LDV) model, ensuring that explanatory power reflected signal extraction rather than trend persistence. The LDV model ($R^2 = 0.848$, Durbin-Watson = 1.899) is treated as the preferred specification given the time-series structure of the data (Appendix C.1).

To handle the high-dimensional, non-linear dependencies between weather friction and human intent, we implemented a Random Forest-based inference core. This allowed the system to perform ensemble prediction across 16 variables to generate demand forecasts under high-uncertainty conditions. Stability was verified through five-fold cross-validation. Feature importance was assessed using permutation importance to avoid bias toward continuous variables. Out-of-sample validity was tested using a chronological split, training on 317 days and evaluating performance on an 80-day hold-out period using $R^2$, Mean Absolute Error (MAE), and Root Mean Square Error (RMSE).

### 3.4 Kansei Information Extraction and Spatial Threshold Identification

To examine affective responses underlying the Under-Vibrancy Paradox, Kansei information science methodologies were applied. A total of 71,623 free-text survey responses were analyzed using rule-based keyword matching to identify affective terms associated with satisfaction and dissatisfaction. Comparative frequency analysis across low-satisfaction (1-2 star) and high-satisfaction (4-5 star) cohorts was used to identify dominant atmospheric drivers of visitor sentiment.

To evaluate daily micro-level carrying capacities, daily satisfaction scores were correlated with relative crowd density. This was conducted to determine whether current visitor volumes at monitored nodes have reached a congestion threshold that degrades the visitor experience.

### 3.5 Cross-Prefectural Signal Analysis and Opportunity Quantification

To assess regional interdependence, a cross-correlation function (CCF) analysis was conducted between tourism survey activity in the adjacent Ishikawa Prefecture and physical AI-camera arrivals in Fukui, identifying lagged spillover effects across administrative borders. Finally, the economic impact of weather deterrence was quantified by establishing a novel, measurable performance indicator: the "Opportunity Gap" metric. On days exhibiting high digital intent coupled with elevated weather severity, the negative residuals of the predictive model were interpreted as suppressed visitation associated with planning friction. These intent-implied visits (865,917) were multiplied by the baseline mean per-capita expenditure of ¥13,811 to yield an estimated annual Opportunity Gap of ¥11.96 billion. This quantification establishes a concrete, metrologically grounded baseline for evidence-based policy making (EBPM) and regional resource allocation.

## 4. Results

The application of the Distributed Human Data Engine (DHDE) produced statistically robust results across predictive modeling, affective text analysis, and economic quantification. The findings demonstrate that integrating digital intent with high-resolution environmental filters enables accurate forecasting of visitor flows, while also revealing structural limitations in prevailing capacity management assumptions for low-density regional destinations.

### 4.1 Predictive Performance and Statistical Robustness

The primary objective of the predictive module was to evaluate whether digital intent signals could reliably forecast physical arrivals at the primary coastal node, **Tojinbo**. The Ordinary Least Squares (OLS) regression model exhibited strong explanatory power, accounting for 81.0 percent of daily visitor variance ($R^2 = 0.810$, Adjusted $R^2 = 0.802$, $p < 0.001$) across the 397-day effective observation period after merging with route search impression (RSI) data (Table 1).

Standardized Beta coefficients were computed to evaluate relative effect sizes. Route search impression (RSI) volume emerged as the dominant behavioral driver ($\beta = +0.456$, $p < 0.001$), while the indicator for weekends and holidays remained the strongest overall predictor ($\beta = +0.547$). This confirms that digital intent functions as the primary psychological driver of visitation, with weather acting as a conditional environmental constraint ($\beta = -0.050$) rather than a competing determinant. The overall model yielded a Cohen's $f^2$ of 4.25, indicating an extremely large effect size.

To validate the system's distributed reliability, the model was deployed across the 4-node network. While the primary coastal node (Node A) achieved $R^2 = 0.482$, predictive performance across other nodes remained robust given their differing noise profiles: Node B (Urban) $R^2 = 0.436$, Node C (Mountain) $R^2 = 0.374$, and Node D (Scenic) $R^2 = 0.168$. To capture non-linear interactions, the system utilized a Random Forest (RF) regressor, which yielded a training $R^2$ of 0.909 and a 5-fold cross-validation $R^2$ of 0.557 (± 0.131). Permutation importance, invariant to inter-feature correlation, confirmed digital search intent and month as the primary drivers, corroborating the OLS beta coefficient rankings through a multicollinearity-robust method. The gap between the training $R^2$ (0.909) and the 5-fold cross-validation $R^2$ (0.557) reflects the sensitivity of Random Forest to randomized fold assignment in temporally ordered data, where future observations can leak into training folds. For time-series validation, randomized k-fold typically produces upward-biased (optimistic) estimates due to temporal leakage across folds; the chronological hold-out evaluation is therefore the more reliable measure of real-world predictive validity. On the 80-day chronological hold-out, the RF attains $R^2 = 0.512$, while the primary OLS specification attains $R^2 = 0.683$, which is the out-of-sample figure reported throughout this paper.

Node D (Scenic Drive: Rainbow Line Parking Lot) requires metrological qualification. The lower node-level explanatory power ($R^2$ = 0.168) reflects a fundamental instrument mismatch intrinsic to this node's ground-truth sensor configuration. The camera at this location is positioned at the vehicle entry gate of the Rainbow Line scenic toll road, using face detection to record occupant faces as vehicles pass through the gate, rather than counting free-flow pedestrians. The digital intent proxy (route search impressions, RSI) captures individual visitor planning intent. The many-to-one relationship between gate face detections and total per-vehicle visitor count introduces systematic, unmodelled variance that cannot be resolved without individual-level physical instrumentation. As a destination-specific scenic drive rather than a multi-modal transit hub, Rainbow Line also exhibits higher sensitivity to day-of-week and weather conditions relative to other nodes, further reducing intent-arrival correspondence. Crucially, Node D contributed approximately 13,142 visits to the total annual proxy gap estimate, representing less than 2% of the 865,917-visit total. The system-level opportunity gap estimate is therefore robust to Node D uncertainty: a sensitivity analysis removing Node D entirely changes the aggregate figure by under 2%.

To address potential time-series autocorrelation, a First-Difference specification ($\Delta y \sim \Delta X$) was estimated. Predictive performance remained high ($R^2$ = 0.708), and the Durbin-Watson statistic improved to 2.524, indicating that model performance was not driven by spurious trend persistence.

Out-of-sample predictive validity was evaluated using a chronological train-test split. Training on 317 days and testing on an 80-day hold-out set, the model achieved an unseen-data predictive performance ($R^2$ = 0.683) with a Mean Absolute Error of 1,793 visitors per day. These results demonstrate that the DHDE operates as a forward-looking predictive system rather than a purely retrospective analytical model (Table 2). The high degree of alignment between the model-predicted demand and the physical ground-truth arrivals is visually illustrated in Figure 2. Notably, the model demonstrated high structural stability when validated as the observational dataset expanded, maintaining an $R^2$ of 0.810 as the dataset expanded through automated daily collection.

Error Decomposition and Uncertainty Propagation: The system-level MAE of 1,793 visits/day reflects three additive uncertainty components. First, sensor measurement error arises from the precision limits of the on-camera YOLO-based human-shape detection algorithm (DNN Network Bullet IP Camera U2SM-B, IP66-rated) operating under variable lighting, crowd density, and adverse weather conditions; this component represents inherent camera-level counting uncertainty that varies across nodes and seasons. Second, data processing error arises in the pipeline converting raw detections into the daily measurand: temporal aggregation of five-minute interval counts into daily windows, the exclusion of 17 documented outage days, and the survey-proxy substitution at Node C. Third, model approximation error represents the residual variation in daily visitor counts not explained by the 16-predictor OLS specification, as reflected by the in-sample $R^2$ = 0.810. The economic opportunity gap estimate (JPY 11.96 billion) inherits these uncertainties proportionally and should be interpreted as an order-of-magnitude proxy figure rather than a precise revenue forecast, consistent with the study's metrological framing.

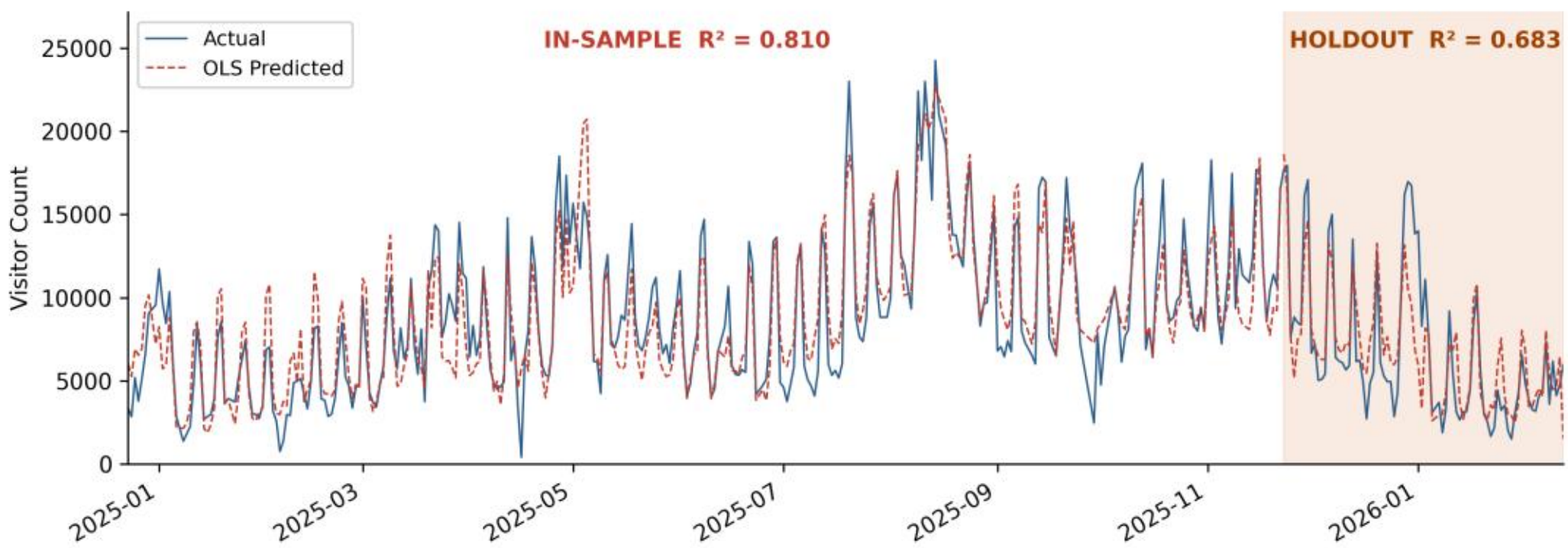

**Figure 2.** Actual daily arrivals versus OLS-predicted arrivals across the 427-day observation window. The shaded region marks the 80-day chronological hold-out period. In-sample $R^2 = 0.810$; hold-out $R^2 = 0.683$ (MAE = 1,793 visitors/day), the primary measure of real-world predictive validity.

## 4.2 Affective Keyword Analysis and the Under-Vibrancy Paradox

Kansei analysis was conducted using 71,623 free-text visitor responses with valid affective content, drawn from the full set of 97,719 standardized survey records. At the macro-regional level, a Spearman rank-order correlation revealed a positive association between visitor satisfaction and monthly visitor density ($r_s = +0.150$, $p = 0.002$). This relationship runs counter to dominant overtourism assumptions and supports the existence of an Under-Vibrancy Paradox, where human presence contributes positively to the visitor atmosphere.

Rule-based analysis of free-text visitor reviews indicates that dissatisfied visitors disproportionately referenced affective terms associated with emptiness, including “lonely,” “deserted,” and “closed shops.” These expressions were observed predominantly in low-satisfaction (one-star and two-star) reviews and were rare or absent in high-satisfaction (five-star) reviews. This qualitative asymmetry (occurring at an 11.5× higher rate) provides empirical support for an Under-Vibrancy Paradox, in which insufficient human presence and inactive commercial environments degrade perceived visitor experience in structurally low-density regions. A formal test of proportions confirms that the prevalence of under-vibrancy keywords is significantly higher in dissatisfied reviews (6.1%) compared to high-satisfaction reviews (~0.5%; $\chi^2 = 514.7$, $p < 0.001$).

Conversely, at the daily micro-level, site-specific carrying capacities exhibited high atmospheric resilience. Daily satisfaction remained essentially constant regardless of density fluctuations at both the natural node and the heritage node ($r = +0.047$, $p = 0.251$ at Eiheiji temple). This indicates that current visitor volumes remain far below the threshold necessary to trigger negative congestion effects, confirming that the region possesses significant latent capacity to absorb rerouted visitor flows without degrading the individual experience.

Qualitative validation at the heritage node (Eiheiji) further supports this resilience. Text mining of 4,504 area-specific responses revealed that while 93.7% of visitors reported high satisfaction (4-5 stars), congestion-related complaints accounted for a mere 0.2% of total feedback. Notably, zero instances of dissatisfaction were attributed to the degradation of the site's spiritual atmosphere by crowds, with the few recorded complaints focused exclusively on peripheral infrastructure such as parking and transit schedules. This confirms that the region's most sensitive cultural landmarks possess the latent capacity to absorb significantly higher visitor volumes.

### 4.3 Quantification of the Economic Opportunity Gap

Discrepancies between predicted arrivals and observed visitor counts enabled direct measurement of planning friction. The model identified 42 days characterized by high digital intent but suppressed physical arrivals due to severe environmental conditions.

Aggregating positive model residuals (days where predicted arrivals exceeded observed counts) across all four monitoring nodes, the DHDE estimates a proxy for unmet visitation of 865,917 visits annually. This figure represents the upper-bound gap between intent-implied demand and realised arrivals on high-friction weather days, not a direct measure of suppressed demand. Using the mean per-capita expenditure of ¥13,811 derived from survey data, the resulting economic leakage associated with planning friction is estimated at ¥11.96 billion (approximately USD 72.6 million (at ¥164.7/$1)) per year (Table 3). To assess the macroeconomic impact of demand recovery, we simulated the effect of the opportunity gap on the prefecture's national tourism ranking (overnight-stay metric). As illustrated in Figure 3, recovering the 865,917 annual intent-implied visits from the four monitored nodes alone closes between 8.4% and 66.2% of the monthly visitor shortfall required to exit the bottom tier of national rankings. While the four nodes provide modest direct rank improvements (e.g. January improving from 46th to 39th), the data suggests a significant scaling effect. Extrapolating the DHDE measurement framework to the prefecture's remaining tourism sites would realistically close the total regional opportunity gap, projecting a peak monthly improvement into the 35th-rank range nationally.

Seasonal sensitivity analysis further showed that the predictive contribution of weather data increased substantially during winter. The incremental explanatory gain ($\Delta R^2$) reached 0.135 in winter compared to 0.021 in summer, yielding a 6.26-fold difference. This indicates that winter tourism economies are disproportionately exposed to weather-associated planning friction.

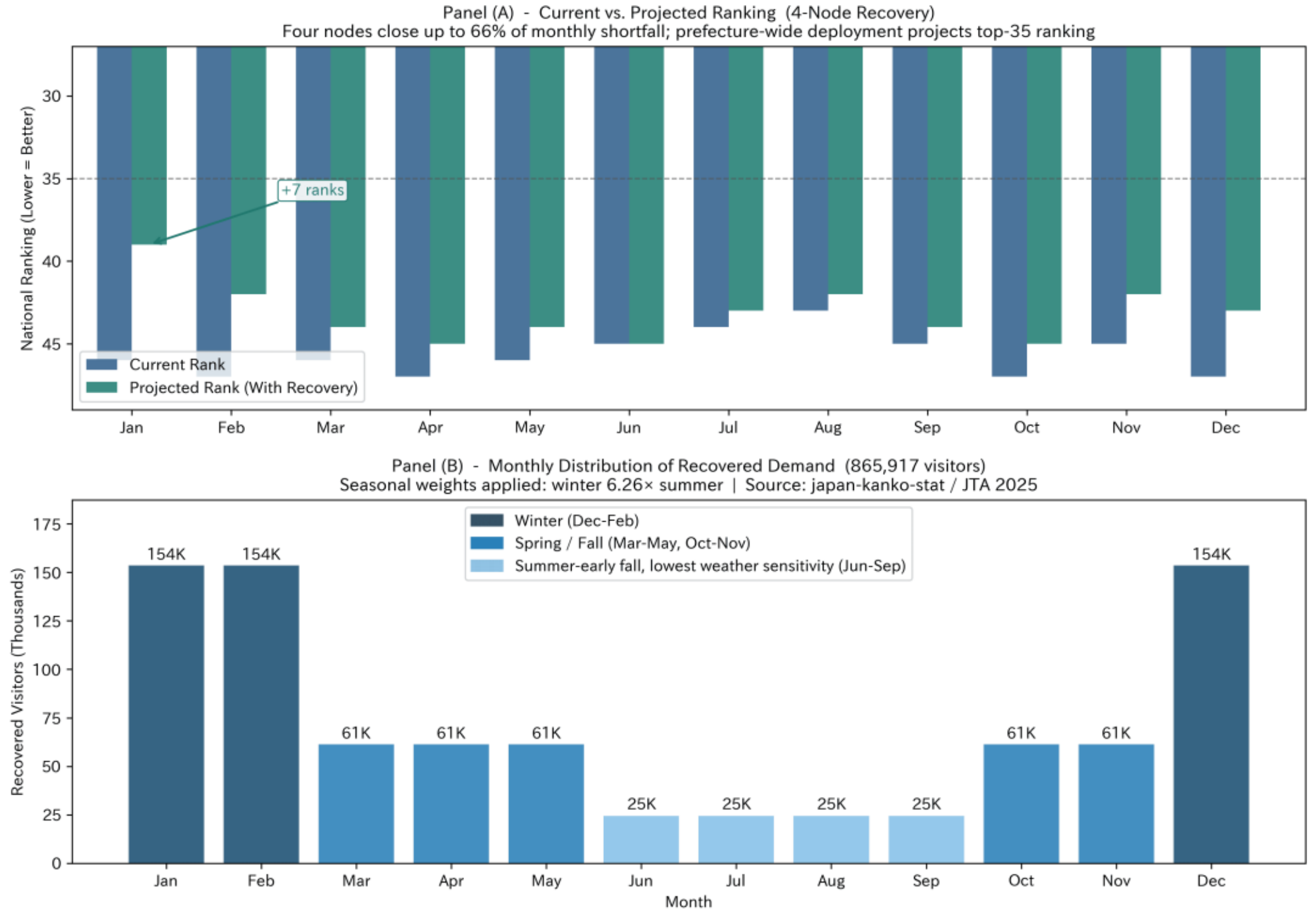

**Figure 3. National ranking recovery potential under AI-driven governance.** Panel (A) compares Fukui Prefecture's current monthly ranking (overnight-stay metric) against the projected improvement assuming visitor recovery at the four monitored nodes. Panel (B) shows the monthly distribution of the 865,917 recovered visitors using seasonal weights. While recovery at these four nodes alone closes up to 66% of the monthly shortfall to the next ranking tier (e.g. January improving from 46th to 39th), prefecture-wide extrapolation of the DHDE framework suggests a realistic path toward a top-35 national ranking.

## 4.4 Spatial Generalization and Inter-Prefectural Flow

To assess regional interdependence, a Cross-Correlation Function analysis was conducted between digital tourism activity in Ishikawa Prefecture and physical arrivals at Fukui's coastal node Tojinbo. A moderate same-day correlation was observed ($r = 0.549$), consistent with regional demand co-movement between Ishikawa and Fukui tourism activity, suggesting cross-prefectural visitor flow patterns that transcend individual administrative boundaries (Figure 4).

To test spatial generalizability, the predictive framework was applied to the **Fukui Station** urban transit hub. Despite the presence of commuter-related noise, the model retained substantial explanatory power ($R^2 = 0.436$). This confirms that digital intent signals can forecast human flow in both destination-focused tourism sites and mixed-use urban infrastructure.

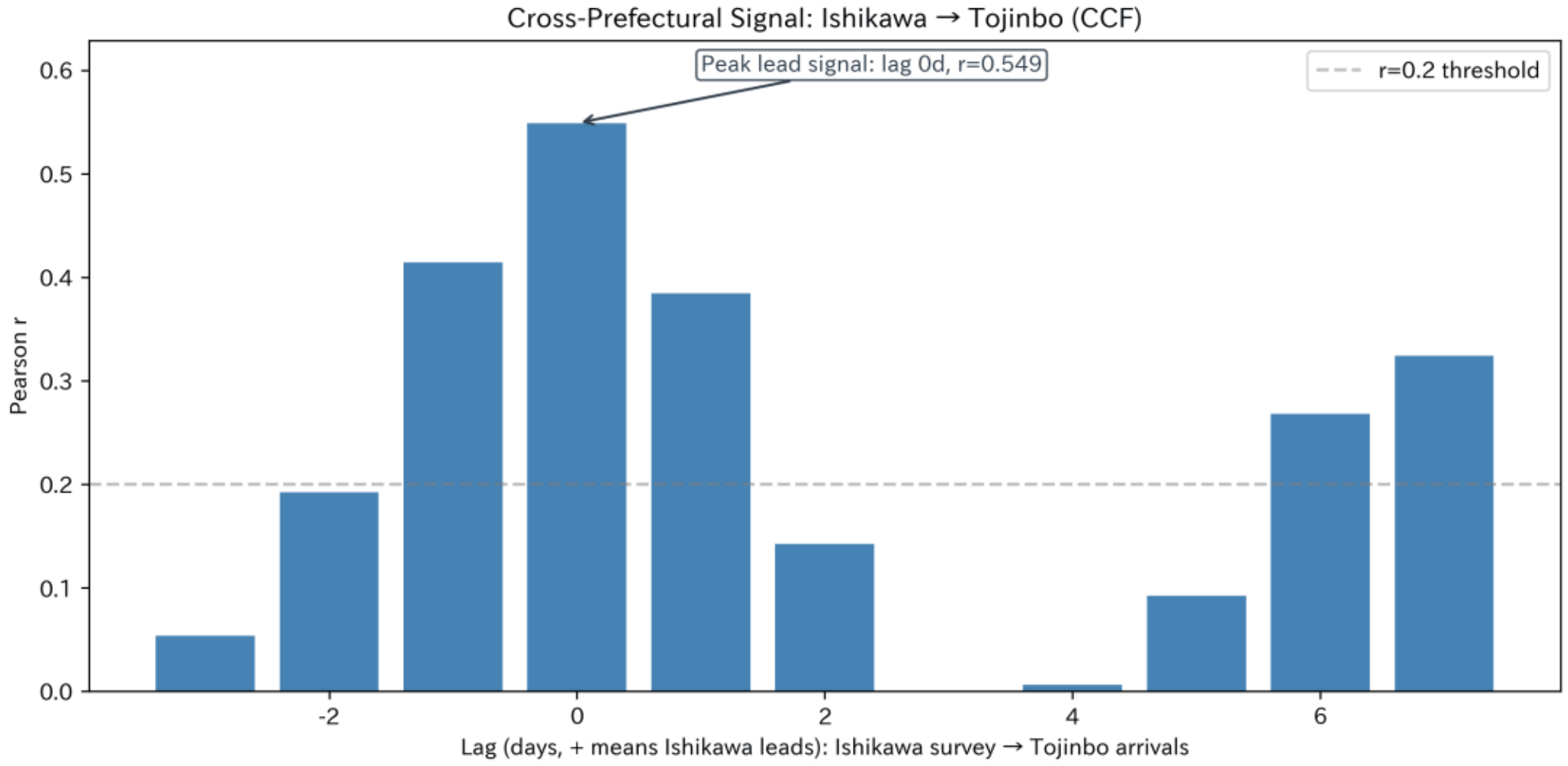

**Figure 4. Cross-prefectural demand spillover signal.** Digital tourism activity in Ishikawa Prefecture shows a statistically significant same-day contemporaneous correlation ($r$ = 0.549 at Lag 0) with physical arrivals in Fukui Prefecture.

## 5. Discussion

The results of this study have significant theoretical and practical implications for the fields of urban informatics, behavioral economics, and regional governance. By applying the Distributed Human Data Engine (DHDE) to a regional tourism economy, we demonstrate that data architectures originally developed for public health crisis management can be recalibrated to address structural economic stagnation.

### 5.1 Theoretical Implications: The Under-Vibrancy Paradox

The primary empirical finding of this research is the *Under-Vibrancy Paradox*. In structurally under-visited regions, crowd density functions as a positive density signal rather than a congestion indicator. The dominant paradigm in smart tourism and urban management literature focuses primarily on overtourism, operating under the assumption that increased visitor density necessarily degrades visitor experience. In contrast, our Kansei-based text analysis and statistical results indicate that for peripheral or structurally under-visited regions, the inverse relationship frequently holds. Visitor satisfaction at natural and commercial nodes is positively correlated with crowd density ($r_s = +0.150, p = 0.002$).

Moreover, dissatisfied visitors are 11.5 times more likely to reference loneliness, empty streets, or closed shops than overcrowding. This pattern suggests that spatial emptiness constitutes a substantial form of psychological planning friction. In such regions, human presence itself functions as a positive environmental attribute. Under these conditions, moderate crowding should be interpreted not as a negative externality, but as a prerequisite for perceived vitality.

This study further refines the concept of carrying capacity by extending this logic to quantitative thresholds. The mathematical flatness of daily satisfaction scores ($r \approx 0.00$) across the monitored nodes confirms that these structurally under-visited regions are operating well below their carrying capacities, providing a wide buffer for the safe redistribution of cross-prefectural visitor flows.

### 5.2 Measurement Application and Instrument Utility

The estimation of a **¥11.96 billion (USD 72.6 million)** annual opportunity gap establishes a clear economic rationale for regional investment in measurement infrastructure. The DHDE's validated outputs demonstrate that fused heterogeneous sensor signals can serve as reliable substitutes for sparse physical ground-truth coverage in peripheral economies. The instrument resolves seasonal measurement asymmetries (winter demand shows 6.26 times greater weather friction than summer) that conventional single-sensor approaches fail to detect. These calibrated estimates provide regional planners with a metrologically grounded evidence base for resource allocation decisions.

## 6. Conclusion

This study advances the science of human mobility measurement by validating the DHDE as a sparse-sensor compensation instrument, specifically a multi-modal fusion architecture that resolves the systematic estimation bias arising from sparse ground-truth coverage in peripheral economies. The instrument's metrological validity is established through cross-node replication across four geographically distinct site types and chronological holdout evaluation. By fusing digital intent signals, high-resolution meteorological data, physical camera counts, and a 97,719-record behavioral survey, we calibrate a validated estimate of regional visitation dynamics.

The findings reframe regional economic stagnation not as a deficiency of resources, but as a quantifiable problem of **planning friction** arising from the misalignment of intent and execution. We empirically identify the Under-Vibrancy Paradox and estimate an annual proxy gap of 865,917 intent-implied visits, corresponding to approximately ¥11.96 billion in foregone regional revenue. Our simulation indicates that recovery at the four monitored nodes closes up to 66% of the shortfall required for rank improvement. These findings demonstrate that the DHDE's metrological outputs, validated against physical ground-truth across four geographically distinct node types, provide a quantifiable, actionable evidence base for regional economic planning.

The primary limitation of this research lies in its geographic focus on a single regional context, which may reflect localized travel norms and climatic conditions. Future research should deploy the DHDE framework across diverse international settings, including mega-events and globally significant heritage sites where visitor density pressures are extreme. By transitioning from static administrative control to dynamic, data-driven governance, policymakers can more effectively balance economic vitality, environmental constraints, and cultural integrity.

Beyond geographic scope, the instrument relies on route search impression (RSI) data as the primary digital intent proxy, introducing a platform dependency: changes to RSI data availability, query logging methodology, or regional coverage would require recalibration. The ground-truth sensor design is also camera-centric, which creates instrument mismatch at vehicle-entry nodes (Section 4.1) and may be unsuitable for privacy-sensitive contexts. Future deployments should evaluate complementary sensing modalities, including multi-frequency smartphone positioning, which has demonstrated meter-level outdoor positioning accuracy [27], and Wi-Fi Channel State Information (CSI) sensing [28], which enables passive visitor detection without optical infrastructure. Integrating such modalities would reduce single-platform dependency and extend DHDE applicability to areas with limited camera coverage or stricter privacy regulation.

Furthermore, while the overarching DHDE system architecture is geography-agnostic, its specific input parameters require localized calibration. The current framework was validated in a region characterized by specific environmental stressors (heavy snowfall, coastal winds) and cultural infrastructure (heritage sites). To apply this methodology to tropical or arid climates, the meteorological friction thresholds must be recalibrated to prioritize variables such as heat indices or monsoon intensity. Similarly, cross-cultural deployment requires linguistic adaptation of the Kansei lexicon to ensure that the psychological nuances of ‘emptiness’ versus ‘tranquility’ are accurately captured in differing demographic contexts.

## 7. Funding

This research did not receive any specific grant from funding agencies in the public, commercial, or not-for-profit sectors.

## 8. Declaration of generative AI and AI-assisted technologies in the manuscript preparation process

The authors used Google Gemini and ChatGPT to assist with language polishing, structural organization, and the mapping of technical terminology to enhance the overall clarity and readability of the manuscript. After using these tools, the authors reviewed, validated, and edited the content as needed and take full responsibility for the integrity and accuracy of the final publication.

## 9. Data Availability Statement

The source code and data for this study are hosted in a public repository. The link has been removed for the peer-review process and will be included in the final publication.

## 10. Declaration of Competing Interests

The authors declare that they have no known competing financial interests or personal relationships that could have appeared to influence the work reported in this study.

## 11. Acknowledgments

The authors would like to express their sincere gratitude to Masanori Satake of the Fukui Tourism Federation and Taisuke Fukuno of Code for FUKUI for their strategic insights into regional tourism management and for establishing the practical foundations of the economic opportunity gap analysis.

We thank Professor Hiroyuki Inoue of the University of Fukui for his expert guidance on Kansei information science and cross-university collaboration. We also acknowledge Professor Makoto Fujiu and Assistant Professor Kosuke Koike of Kanazawa University, and Professor Yoichi Kanayama of Toyama

University for their contributions to the data integration framework via the Hokuriku Inbound Tourism DX and Data Consortium.

We thank Bunto Hanyuda of the Ishikawa Prefectural Government for his technical stewardship of the regional open data ecosystem. We further acknowledge the Fukui Prefectural Government for provision of the AI-camera people-flow datasets, and the Japan Meteorological Agency (JMA) for the high-resolution meteorological data underpinning the environmental filtering models.

Finally, the authors thank the local merchants and tourism associations across the Hokuriku region for their contribution to the open datasets used for this study, in addition to the tens of thousands of tourists who contributed to the survey datasets.

Open Access Funding for this article was provided by the University of Fukui.

## References

[1] O. Oklevik, S. Gössling, C.M. Hall, J.K. Steen Jacobsen, H.I. Grøger, S. McCabe, Overtourism, optimisation, and destination performance indicators: A case study of activities in Fjord Norway, J. Sustain. Tour. 27 (2019) 1804–1824. https://doi.org/10.1080/09669582.2018.1533020

[2] H. Seraphin, P. Sheeran, M. Pilato, Over-tourism and the fall of Venice as a destination, J. Destin. Mark. Manag. 9 (2018) 374–376. https://doi.org/10.1016/j.jdmm.2018.01.011

[3] A. Khanzada, F.H. Cheema, T. Takemoto, Improving clinical trial enrollment for smartphone-based AI data collection: a methodological analysis of nudge-based interventions, J. Behav. Econ. Policy 9 (1) (2025) 41-50.

[4] H. Inoue, Y. Mitani, The relationship between university students' emotional attachment to Fukui Prefecture and their employment locations through Kansei analysis, Nat. Environ. Sea Japan Dist. 32 (2025) 51–62. http://hdl.handle.net/10098/0002000666

[5] D. Buhalis, A. Amaranggana, Smart tourism destinations: Enhancing tourism experience through personalisation of services, in: I. Tussyadiah, A. Inversini (Eds.), Information and Communication Technologies in Tourism 2015, Springer, 2015, pp. 377–389. https://doi.org/10.1007/978-3-319-14343-9_28

[6] U. Gretzel, M. Sigala, Z. Xiang, C. Koo, Smart tourism: Foundations and developments, Electron. Mark. 25 (2015) 179–188. https://doi.org/10.1007/s12525-015-0196-8

[7] R. Dodds, R. Butler (Eds.), Overtourism: Issues, Realities and Solutions, De Gruyter Oldenbourg, 2019. https://doi.org/10.1515/9783110607369

[8] C. Milano, J.M. Cheer, M. Novelli (Eds.), Overtourism: Excesses, Discontents and Measures in Travel and Tourism, CABI, 2019. https://doi.org/10.1079/9781786399823.0000

[9] S. Pike, Destination Marketing: Essentials, 2nd ed., Routledge, 2018. https://doi.org/10.4324/9781003032205

[10] J.A. Burke, D. Estrin, M. Hansen, A. Parker, N. Ramanathan, S. Reddy, M.B. Srivastava, Participatory sensing, UCLA Center for Embedded Network Sensing, 2006. https://escholarship.org/uc/item/19h777qd

[11] N.D. Lane, E. Miluzzo, H. Lu, D. Peebles, T. Choudhury, A.T. Campbell, A survey of mobile phone sensing, IEEE Commun. Mag. 48 (2010) 140–150. https://doi.org/10.1109/MCOM.2010.5560598

[12] N. Kankanamge, T. Yigitcanlar, A. Goonetilleke, M. Kamruzzaman, Can volunteer crowdsourcing reduce disaster risk? A systematic review of the literature, Int. J. Disaster Risk Reduct. 35 (2019) 101097. https://doi.org/10.1016/j.ijdrr.2019.101097

[13] A.J. Meijer, M. Thaens, Urban technological innovation: Developing and testing a socio-technical framework for studying smart city projects, Urban Aff. Rev. 54 (2018) 363–387. https://doi.org/10.1177/1078087416670274

[14] X. Yan, Z. Jiang, P. Luo, H. Wu, A. Dong, F. Mao, Z. Wang, H. Liu, Y. Yao, A multimodal data fusion model for accurate and interpretable urban land use mapping with uncertainty analysis, Int. J. Appl. Earth Obs. Geoinf. 129 (2024) 103805. https://doi.org/10.1016/j.jag.2024.103805

[15] H. Choi, H. Varian, Predicting the present with Google Trends, Econ. Rec. 88 (2012) 2–9. https://doi.org/10.1111/j.1475-4932.2012.00809.x

[16] P.F. Bangwayo-Skeete, R.W. Skeete, Can Google data improve the forecasting performance of tourist arrivals? Mixed-data sampling approach, Tour. Manag. 46 (2015) 454–464. https://doi.org/10.1016/j.tourman.2014.07.014

[17] X. Li, B. Pan, R. Law, X. Huang, Forecasting tourism demand with composite search index, Tour. Manag. 59 (2017) 57–66. https://doi.org/10.1016/j.tourman.2016.07.005

[18] H. Song, R.T.R. Qiu, J. Park, A review of research on tourism demand forecasting, Ann. Tour. Res. 75 (2019) 338–362. https://doi.org/10.1016/j.annals.2018.12.001

[19] J. Day, N. Chin, S. Sydnor, K. Cherkauer, Weather, climate and tourism performance: A quantitative analysis, Tour. Manag. Perspect. 5 (2013) 51–56. https://doi.org/10.1016/j.tmp.2012.11.001

[20] R.H. Thaler, C.R. Sunstein, Nudge: Improving Decisions about Health, Wealth, and Happiness, Yale University Press, 2008.

[21] Behavioural Insights Team, EAST: Four simple ways to apply behavioural insights, Annu. Rev. Policy Des. 5 (2017) 1–53. https://ojs.unbc.ca/index.php/design/article/view/1658

[22] M. Weinmann, C. Schneider, J. vom Brocke, Digital nudging, Bus. Inf. Syst. Eng. 58 (2016) 433–436. https://doi.org/10.1007/s12599-016-0453-1

[23] J. Danaher, The threat of algocracy: Reality, resistance and accommodation, Philos. Technol. 29 (2016) 245–268. https://doi.org/10.1007/s13347-015-0211-1

[24] Y. Zheng, L. Capra, O. Wolfson, H. Yang, Urban computing: Concepts, methodologies, and applications, ACM Trans. Intell. Syst. Technol. 5 (2014) Article 38. https://doi.org/10.1145/2629592

[25] M. Nagamachi, Kansei engineering: A new ergonomic consumer-oriented technology for product development, Int. J. Ind. Ergon. 15 (1995) 3–11. https://doi.org/10.1016/0169-8141(94)00052-5

[26] S. Schütte, J. Eklund, J. Axelsson, M. Nagamachi, Concepts, methods and tools in Kansei engineering, Theor. Issues Ergon. Sci. 5 (2004) 214–231. https://doi.org/10.1080/1463922021000049980

[27] B. Wang, Z. Han, C. Liu, Y. Wu, Multi-frequency smartphone positioning performance evaluation: insights into A-GNSS PPP-B2b services and beyond, Satell. Navig. 5 (2024) 35. https://doi.org/10.1186/s43020-024-00146-5

[28] F. Miao, Y. Huang, Z. Lu, T. Ohtsuki, G. Gui, H. Sari, Wi-Fi Sensing Techniques for Human Activity Recognition: Brief Survey, Potential Challenges, and Research Directions, ACM Comput. Surv. 57 (5) (2025). https://doi.org/10.1145/3705893

## Tables

**Table 1. Ordinary Least Squares (OLS) regression results predicting physical arrivals at the primary coastal node (Tojinbo).** The model integrates digital intent signals, high-resolution micro-climate data, and temporal interaction terms, successfully explaining 81.0% of the daily visitor variance.

| Variable | Coefficient | *p*-value |
| --- | --- | --- |
| const | −722.33 | 0.4482 |
| directions | +0.9174 *** | 0.0000 |
| directions_lag1 | +0.3048 * | 0.0353 |
| directions_lag2 | −0.0230 | 0.8718 |
| directions_lag3 | −0.2662 * | 0.0482 |
| directions_roll7 | +0.3872 * | 0.0435 |

| precip | −24.34 | 0.1709 |
|---|---|---|
| temp | −29.33 | 0.0707 |
| sun | +1,294.20 ** | 0.0057 |
| wind | −12.65 | 0.8689 |
| precip_lag1 | −4.16 | 0.7126 |
| is_weekend_or_holiday | +5,374.23 *** | 0.0000 |
| weather_severity | −283.40 | 0.2644 |
| dow_mean_count | −0.0585 | 0.5923 |
| weekend_x_severity | −1,194.52 *** | 0.0000 |
| weekend_x_intent | −0.2271 * | 0.0367 |
| month | +409.47 *** | 0.0000 |
| **$R^2$** | **0.8096** | |
| **Adj. $R^2$** | **0.8016** | |
| **N** | **397** | |

Note: * $p < 0.05$, ** $p < 0.01$, *** $p < 0.001$.

**Table 2. Evaluation of statistical rigor, effect size, and out-of-sample predictive validity.** Panel A ranks standardized beta coefficients ($\beta$), mathematically confirming that digital search intent is the primary behavioral driver, while weather acts as a secondary environmental filter. Panel B demonstrates

an exceptionally large global effect size (Cohen's $f^2 = 4.25$). Panel C validates out-of-sample predictive validity ($R^2 = 0.683$) using an 80-day unseen hold-out set, demonstrating the model's forward-looking predictive utility for regional planning.

| **Feature / Metric** | **Value** | **\|$\beta$\| Rank / Note** |
|---|---|---|
| **Panel A: Standardised Coefficients ($\beta$)** | | |
| is_weekend_or_holiday | +0.5471 | 1 |
| directions | +0.4559 | 2 |
| month | +0.3307 | 3 |
| weekend_x_intent | −0.1658 | 4 |
| weekend_x_severity | −0.1592 | 5 |
| directions_roll7 | +0.1584 | 6 |
| directions_lag1 | +0.1510 | 7 |
| directions_lag3 | −0.1326 | 8 |
| sun | +0.0856 | 9 |
| temp | −0.0601 | 10 |
| precip | −0.0507 | 11 |
| weather_severity | −0.0503 | 12 |
| dow_mean_count | −0.0291 | 13 |
| directions_lag2 | −0.0114 | 14 |
| precip_lag1 | −0.0087 | 15 |
| wind | −0.0043 | 16 |
| **Panel B: Global Effect Size** | | |
| Cohen's $f^2$ | 4.2519 | large (≥ 0.35) |
| **Panel C: Out-of-Sample Predictive Validity** | | |
| Training N | 317 | |
| Hold-out N | 80 | |
| Hold-out MAE | 1,793.2 | visitors/day |
| Hold-out RMSE | 2,461.0 | visitors/day |
| Hold-out $R^2$ | 0.6834 | |

**Table 3. Summary of key economic and spatial governance metrics.** This table quantifies the regional "Opportunity Gap," representing approximately ¥11.96 billion in leaked tourism revenue across the monitored nodes. It highlights the severe susceptibility of winter tourism to weather friction, which is 6.26 times higher than in summer, and confirms a statistically significant cross-prefectural demand spillover from Ishikawa Prefecture ($r = 0.549$ at lag 0).

| **Metric** | **Value** |
|---|---|

| Lost Visitors (single-node) | 85,522 |
|---|---|
| Lost Visitors (4-node) | 865,917 |
| Opportunity Gap Value (¥) | ¥11,959,183,083 |
| Ishikawa CCF *r* | 0.5490 |
| Best lag (days) | 0 |
| Weather Sensitivity Ratio (W/S) | 6.26× |

# Appendices

## A. Data Dictionary and Data Provenance

This appendix delineates the data provenance for the raw records processed by the Distributed Human Data Engine (DHDE) pipeline. It provides the mapping of Japanese source-field terminology to the corresponding English variable names utilized throughout the analytical framework.

### A.1 AI-Camera People-Flow (Primary Ground Truth)

Camera data were sourced as 5-minute interval CSV files from the Fukui Prefecture AI camera people-flow sensor network (DNN Network Bullet IP Camera U2SM-B; IP66-rated; on-camera YOLO-based inference), maintained within the *fukui-kanko-people-flow-data* repository. These intervals were subsequently aggregated to yield daily visitor totals.

**Daily Aggregation Formula:** The daily visitor arrival metric is computed as the sum of all 5-minute intervals $i$ within a 24-hour period $d$:

$$count_day = \Sigma count_(5min(i))$$

Days with a zero count were excluded from the analysis as they indicated documented sensor outages.

| Source CSV Column | Pipeline Variable | Unit | Notes |
|---|---|---|---|
| aggregate from | — | Timestamp | Interval start; utilized for deduplication. |
| total count | count | Persons/day | Summation across all 5-minute intervals. |

**Geographic Node-to-Sensor Mapping**

| Node | Designation | Environment | Camera Source Path |
|---|---|---|---|
| A | Tojinbo / Mikuni | Coastal (North) | tojinbo-shotaro/Person/**/*.csv |
| B | Fukui Station | Urban Transit (Central) | fukui-station-east-entrance/Person/**/*.csv |
| C | Katsuyama / Dinosaur Museum | Mountainous (East) | Survey-response proxy (see note) |
| D | Rainbow Line / Wakasa | Scenic Drive (South) | rainbow-line-parking-lot-1-gate/Face/**/*.csv |

**Note on Measurement Limitations:** Node C utilized a survey-response proxy to estimate visitor counts due to an absence of camera coverage during the observation period. This proxy was validated by correlating daily survey volumes with ground-truth camera counts at the primary node ($r = 0.564$, $p < 0.001$), confirming that survey response frequency is a reliable indicator of physical density in this regional context. This proxy transfer assumes that the metrological relationship between survey participation rate and physical arrivals is spatially invariant across the prefecture's core tourist attractions, an assumption supported by the survey's exclusive focus on domestic Japanese visitors, whose behavioral propensity to complete surveys is unlikely to vary systematically by attraction type. Node D relied on facial-detection counts from a primary parking gate camera; because this represents a sub-sample of total vehicle occupancy rather than a total site count, it exhibits higher instrumental noise ($R^2 = 0.168$) compared to other nodes.

### A.2 Japan Meteorological Agency (JMA) Meteorological Data

Hourly meteorological observations were obtained from four JMA automated weather stations (Mikuni, Fukui City, Katsuyama, and Mihama). Hourly records were aggregated into daily statistics: precipitation (precip) was summed; all other variables were averaged to derive daily means.

| Source CSV Column | Pipeline Variable | Aggregation | Unit |
|---|---|---|---|
| timestamp | date | Normalized to midnight | — |
| temp_c | temp | Mean | °C |
| precip_1h_mm | precip | Sum | mm/day |
| sun_1h_h | sun | Mean | Hours/day |
| wind_speed_ms | wind | Mean | m/s |
| snow_depth_cm | snow_depth | Mean | cm (Node B only) |
| humidity_pct | humidity | Mean | % |

### A.3 Route Search Impressions (Digital Intent Signal)

Daily route search impression (RSI) counts were obtained from the publicly available *fukui-kanko-trend-report* open data repository (Code for Fukui), which provides pre-aggregated impression counts from online maps and web search tools. The repository is available at https://github.com/code4fukui/fukui-kanko-trend-report.

| Source CSV Column | Pipeline Variable | Unit | Notes |
|---|---|---|---|
| directions | directions | Searches/day | Count of route-to-location searches; primary digital intent proxy. |

### A.4 Hokuriku Visitor Survey

Two complementary survey datasets are used by the pipeline; they serve distinct analytical purposes. Dataset 2 is a standardized subset of Dataset 1, augmented with Ishikawa and Toyama records.

**Dataset 1: Fukui-specific raw survey (fukui-kanko-survey/all.csv):** Contains 90,350 individual responses collected at Fukui Prefecture tourism sites. Used exclusively for spending analysis (県内消費額 → ¥ midpoints). Note: 都道府県 in this file records the visitor's home prefecture, not the collection site.

**Dataset 2: Hokuriku three-prefecture merged survey (opendata/output_merge/merged_survey_*.csv):** Contains 97,719 standardized responses drawn from four survey exports spanning April 2023 to March 2026. The survey was administered at tourism sites within the three Hokuriku prefectures: Fukui (71,623 responses), Ishikawa (20,990), and Toyama (5,106). This dataset supplies Net Promoter Scores (NPS), site-prefecture data, and the 71,623-response Fukui free-text corpus used for under-vibrancy text mining (Appendix D).

**Relationship between datasets:** The 71,623 Fukui-site responses in Dataset 2 are a standardized subset of Dataset 1's 90,350 responses (those records that passed the merged export's completeness criteria). The remaining ~18,727 Dataset 1 records are available only in all.csv. Dataset 2 adds Ishikawa and Toyama responses not present in Dataset 1.

**Dataset 2 Column Mapping (merged_survey_*.csv)**

| Source CSV Column | Pipeline Variable | Data Type | Notes |
|---|---|---|---|
| 対象県（富山/石川/福井） (col 0) | prefecture | String | Survey collection site's prefecture (富山, 石川, or 福井). |
| アンケート回答日 (col 1) | survey_date | Date | Date of the recorded visit. |

| Source CSV Column | Pipeline Variable | Data Type | Notes |
|---|---|---|---|
| 満足度（旅行全体） | satisfaction | Integer (1-5) | Overall trip satisfaction. |
| おすすめ度 | nps_raw | Integer (0-10) | Raw Net Promoter Score. |
| 満足度（商品・サービス） | satisfaction_service | Integer (1-5) | Service satisfaction. |
| 満足度理由 | reason | String | Free-text: reason for visiting. |
| 不便 (partial match) | inconvenience | String | Free-text: reported inconveniences. |
| 自由意見 (partial match) | freetext | String | General free-text commentary. |
| 回答場所 | location | String | Specific site of survey collection. |

**Satisfaction Scale Mapping**

| Japanese Label | English Translation | Integer Score |
|---|---|---|
| とても不満 | Very dissatisfied | 1 |
| 不満 | Dissatisfied | 2 |
| どちらでもない | Neutral | 3 |
| 満足 | Satisfied | 4 |
| とても満足 | Very satisfied | 5 |

## B. Feature Engineering and Derived Variables

### B.1 Temporal Intent Features

To model the temporal gap between planning and execution, the following features were engineered from the route search impression (RSI) volume (intent):

1. **Lagged Intent:** *intent_lag_k = intent_(t-k), where k ∈ {1, 2, 3} days*
2. **Rolling Mean:** *intent_roll_7 = (1/7) × Σ intent_(t-i) for i = 0 to 6*
3. **Weekend Interaction:** *weekend_x_intent = intent_t × is_weekend_or_holiday*

### B.2 Weather Severity Algorithm

A composite ordinal index (0–3) was developed to standardize meteorological friction across different topographies:

| Severity Score | Category | Precipitation Threshold | Wind Speed Threshold |
|---|---|---|---|
| 0 | Fine | 0 mm/day | ≤ 8 m/s |
| 1 | Marginal | > 0 and ≤ 10 mm/day | ≤ 8 m/s |
| 2 | Hostile | > 10 mm/day | ≤ 8 m/s |
| 3 | Severe | > 10 mm/day | > 8 m/s |

### B.3 Full Engineered Feature Set

| Engineered Feature | Derivation | Description |
|---|---|---|
| directions | Raw | Same-day route search impression volume. |
| directions_lag1 to lag3 | Time-shifted (1-3 days) | Lagged digital intent to capture planning horizons. |
| directions_roll7 | 7-day rolling mean | Smoothed trend of digital intent. |
| precip | JMA daily sum | Total daily precipitation (mm). |
| temp | JMA daily mean | Average daily temperature (°C). |
| sun | JMA daily mean | Average daily sunshine duration. |
| wind | JMA daily mean | Average daily wind speed (m/s). |
| precip_lag1 | Time-shifted (1 day) | Prior-day precipitation accumulation. |
| is_weekend_or_holiday | Calendar logic | Binary indicator (1 = weekend or national holiday). |
| weather_severity | Threshold-based | Ordinal scale (0-3) reflecting meteorological hostility. |
| dow_mean_count | Grouped mean | Historical average visitor count by day of week. |
| weekend_x_severity | Interaction term | Peak calendar days × adverse weather. |
| weekend_x_intent | Interaction term | Amplification of digital intent on peak calendar days. |
| month | Extracted from date | Calendar month (1-12) for macroeconomic seasonality. |

## C. Statistical Robustness and Diagnostics

### C.1 Augmented Dickey-Fuller (ADF) Stationarity Tests

ADF tests were applied to the two primary time series over the 427-day analysis window (December 20, 2024 to March 10, 2026), using statsmodels.tsa.stattools.adfuller with automatic lag selection based on the Akaike Information Criterion (AIC).

| Time Series | ADF Statistic | p-value | Optimal Lag (AIC) | Stationarity Decision |
|---|---|---|---|---|
| count (Daily visitor arrivals) | -2.916 | 0.0435 | 14 | Stationary (Reject $H_0$ at 5%) |
| directions (route search impressions) | -2.480 | 0.1204 | 15 | Non-Stationary (Fail to reject $H_0$) |

**Interpretation:** This mixed I(0)/I(1) structure accounts for the positive autocorrelation observed in the baseline OLS residuals (Durbin-Watson = 1.005) and necessitates the robust specifications detailed in C.3. Because the dependent variable is stationary while key predictors exhibit mild trend behavior, the Lagged Dependent Variable (LDV) model is adopted as the preferred specification ($R^2 = 0.848$, Durbin-Watson = 1.899), supplemented by Newey-West standard errors and a first-difference specification as additional robustness checks.

### C.2 Variance Inflation Factors (VIF)

VIF analysis was performed on all 16 model features using statsmodels.stats.outliers_influence.variance_inflation_factor to assess multicollinearity.

| Feature | VIF | Classification |
|---|---|---|
| is_weekend_or_holiday | 15.0 | High (calendar baseline) |
| weekend_x_intent | 12.5 | High (interaction term) |
| directions_roll7 | 12.2 | High (autocorrelated trend) |
| directions | 9.8 | Moderate |
| weather_severity | 4.0 | Low |
| precip | 2.7 | Low |
| temp | 2.2 | Low |

**Interpretation:** Elevated VIF values for intent-related features inflate standard errors and limit the precision of individual coefficient estimates for those variables; collinearity should therefore be considered when interpreting the magnitude of specific betas. However, model-agnostic permutation importance independently confirms directions as the dominant predictor, and Newey-West standard error corrections maintain valid t-statistics under heteroskedasticity and autocorrelation, supporting the overall model's inferential stability.

### C.3 Model Specification Comparison

| Model Specification | $R^2$ | Durbin-Watson | Diagnostic Status |
|---|---|---|---|
| Baseline OLS | 0.8096 | 1.005 | Positive autocorrelation detected |
| OLS + Newey-West HAC | 0.8096 | 1.005 | Robust standard errors applied |
| First-Difference (Δy ~ ΔX) | 0.7083 | 2.525 | Autocorrelation removed |
| Lagged Dependent Variable (LDV) | 0.8485 | 1.899 | Preferred specification |

The LDV model is defined as:

$$count_t = \beta_0 + \beta_1\ count_(t-1) + \beta_k\ X_t + \varepsilon_t$$

**Interpretation:** The LDV specification successfully absorbed the autoregressive component of visitor arrivals, yielding an optimal Durbin-Watson statistic of 1.899. The first-difference model ($R^2 = 0.7083$) confirms that the predictive signal remains robust after removing trend persistence.

**Value of Weather Integration:** Excluding the five JMA meteorological features reduced the baseline OLS $R^2$ from 0.8096 to 0.7537 ($\Delta R^2 = 0.0559$). Seasonal analysis indicates weather sensitivity is 6.26 times more pronounced during winter months ($\Delta R^2 = 0.1349$) than in summer months ($\Delta R^2 = 0.0215$).

## D. Text Mining Methodology and Under-Vibrancy Lexicon

### D.1 Preprocessing Pipeline

A total of 71,623 free-text survey responses from Fukui tourism sites were processed using the following steps:

1. UTF-8 normalization of Japanese characters
2. Removal of null or blank entries
3. Concatenation of satisfaction reason (満足度理由), inconvenience (不便), and general opinion (自由意見) fields
4. Trimming of leading/trailing whitespace
5. Normalization of Roman and Katakana characters to half-width equivalents

### D.2 Keyword Detection Logic

Keyword detection used exact root-form substring matching (str.contains) across concatenated free-text fields. A morphological tokenizer was not applied; root-form matching successfully captures conjugated variants (e.g., matching 「寂し」 captures 「寂しい」 and 「寂しかった」).

### D.3 Under-Vibrancy Keyword Lexicon

| # | Japanese Keyword | Romanization | Category | English Gloss |
|---|---|---|---|---|
| 1 | 静か | shizuka | Atmosphere | Quiet / Silent |
| 2 | 寂し | sabishi | Atmosphere | Lonely / Desolate |
| 3 | さびし | sabishi | Atmosphere | Lonely (hiragana) |
| 4 | さみし | samishi | Atmosphere | Lonely (phonetic variant) |
| 5 | 人が少な | hito ga suku-na | Density | Few people around |
| 6 | 人がいな | hito ga i-na | Density | Nobody around |
| 7 | 活気 | kakki | Atmosphere | Vitality (absence of) |
| 8 | 賑わ | nigiwai | Atmosphere | Lively (absence of) |
| 9 | にぎわ | nigiwai | Atmosphere | Lively (hiragana) |
| 10 | 閑散 | kansan | Atmosphere | Deserted / Sparse |
| 11 | 寂れ | sabie | Decline | Run-down |
| 12 | さびれ | sabie | Decline | Run-down (hiragana) |
| 13 | 閉まっ | shimatte | Commerce | Closed facilities |
| 14 | 店がな | mise ga na | Commerce | No shops present |
| 15 | 営業し | eigyo shi | Commerce | Operating (negative constructions) |
| 16 | 何もな | nani mo na | Experience | Nothing to do |
| 17 | つまらな | tsumarana | Experience | Boring / Dull |
| 18 | 退屈 | taikutsu | Experience | Boredom |
| 19 | 物足りな | monotari-na | Experience | Unsatisfying |
| 20 | 盛り上が | moriagari | Atmosphere | Excitement (absence of) |
| 21 | 人通り | hitodori | Density | Foot traffic |

### D.4 Sentiment Analysis Summary

| Metric | Value |
|---|---|
| Total low-satisfaction (1-2 star) responses analyzed | 1,066 |
| Under-vibrancy keyword occurrences in low-satisfaction responses | 65 |

| Metric | Value |
|---|---|
| Prevalence rate in low-satisfaction responses | 6.1% |
| Prevalence rate in high-satisfaction (4-5 star) responses | ~0.5% |
| Comparative ratio (low vs. high satisfaction) | 11.5× |
| Chi-square statistic | 514.7 |
| p-value | < 0.001 |

## E. Meteorological Station Metadata

Stations were selected based on proximity to the designated analytical nodes. Three stations are from the Automated Meteorological Data Acquisition System (AMeDAS); one is a main observatory.

| Station | Type | block_no | Latitude (N) | Longitude (E) | Elevation | Assigned Node | Distance |
|---|---|---|---|---|---|---|---|
| Mikuni (三国) | AMeDAS | 1071 | 36°13.3’ | 136°08.9’ | 5 m | Node A (Tojinbo) | ~3 km |
| Fukui City (福井) | Main Obs. | 47616 | 36°03.4’ | 136°13.3’ | 9 m | Node B (Fukui Station) | ~1 km |
| Katsuyama (勝山) | AMeDAS | 1226 | 36°03.6’ | 136°30.0’ | 160 m | Node C (Katsuyama) | ~2 km |
| Mihama (美浜) | AMeDAS | 1010 | 35°35.8’ | 135°57.3’ | 5 m | Node D (Rainbow Line) | ~5 km |

**Climatological context by station:**

- **Mikuni:** Coastal Sea of Japan exposure; primary friction from sea-effect wind and precipitation. Snow depth is not recorded at this AMeDAS station.
- **Fukui City:** Urban lowland basin (9 m elevation). Only station providing validated snow depth data ($\beta$ = -0.0056 standardized), making it the most snow-sensitive node.
- **Katsuyama:** Highland mountain basin (160 m elevation); highest-elevation monitored zone. Snow depth not recorded by this AMeDAS instrument; notable diurnal temperature variation (±8–12°C).
- **Mihama:** Coastal Wakasa Bay; oceanic thermal moderation with moderate rainfall. Node D achieves a weather lift of $\Delta R^2$ = +0.039 with Mihama data.

## F. Reproducibility

All data processing, Kansei text mining, and econometric modeling were implemented in Python 3.12.3 using the following core libraries:

| **Library** | **Version** | **Purpose** |
|---|---|---|
| Pandas | 3.0.1 | Data harmonization and cleaning |
| Statsmodels | 0.14.6 | OLS regression, ADF tests, Durbin-Watson diagnostics |
| Scikit-learn | 1.8.0 | Random Forest Regressor and Permutation Importance |
| numpy | 2.4.2 | Numerical computation |
| jpholiday | 1.0.3 | Japanese national holiday calendar logic |